\documentclass[12pt,twoside]{ord}

\usepackage{amssymb}
\usepackage{amsmath}     
\usepackage{algorithmic} 
\usepackage[T1]{fontenc}
\usepackage[]{algorithm2e}
\usepackage{lineno}	

\usepackage{bbm}
\usepackage{tikz}
\usetikzlibrary{positioning}
\usetikzlibrary{arrows.meta}

\newcommand{\chg}[1]{\textcolor{black}{#1}}
\newcommand{\m}[1]{\texttt{#1}}

\ORDtitle{Combining predictive distributions of electricity prices: Does minimizing the CRPS lead to optimal decisions in day-ahead bidding?}
\ORDshortafill{W. Nitka and R. Weron}
\ORDshorttitle{Combining predictive distributions of electricity prices: Does minimizing the CRPS lead to optimal decisions?}
\ORDaffilone{Department of Operations Research and Business Intelligence, Wroc{\l}aw University of Science and Technology, Wroc{\l}aw, Poland}
\ORDaffiltwo{}
\ORDaffilthree{}
\ORDaffilfour{}
\ORDauthorone{Weronika~Nitka}{*1}{0000-0002-3360-4260}{}
\ORDauthortwo{Rafa{\l}~Weron}{1}{0000-0003-1619-5239}{}
\ORDauthorthree{}{2}{000-000200-3}{}
\ORDauthorfour{}{3}{000-000200-4}{}
\ORDauthorfive{}{3}{000-000200-5}{}
\ORDcorrespoding{weronika.nitka@pwr.edu.pl}

\ORDnumber{3}
\ORDvolume{33}
\ORDdoi{DOI: draft}
\ORDyear{2023}
\ORDReceived{xxxxx~xxxxx}
\ORDRevised{xxxxx~xxxxx}
\ORDAccepted{xxxxx~xxxxx}
\ORDPublished{xxxxx~xxxxx}
\ORDFirstpagenum{1}

\begin{document}

\maketitle

\begin{abstract}
Probabilistic price forecasting has recently gained attention in power trading because decisions based on such predictions can yield significantly higher profits than those made with point forecasts alone. At the same time, methods are being developed to combine predictive distributions, since no model is perfect and averaging generally improves forecasting performance.
In this article we address the question of whether using CRPS learning, a novel weighting technique minimizing the continuous ranked probability score (CRPS), leads to optimal decisions in day-ahead bidding. To this end, we conduct an empirical study using hourly day-ahead electricity prices from the German EPEX market. We find that increasing the diversity of an ensemble can have a positive impact on accuracy. At the same time, the higher computational cost of using CRPS learning compared to an equal-weighted aggregation of distributions is not offset by higher profits, despite significantly more accurate predictions.
\end{abstract}


\ORDKeywords{decision support, day-ahead electricity bidding, predictive distribution, combining forecasts, CRPS learning} 


\section{Introduction}
\label{sec:intro}

In order to mitigate risks or increase profits from trading in day-ahead power markets, market participants use data-driven decision support techniques \chg{\cite{janczura_arxgarch_2023,maciejowska_portfolio_2022,maciejowska_enhancing_2021,vogler_eventbased_2022}}. For years, these have relied on point forecasts of the major variables of interest: loads (or demand for electricity), generation from renewable energy sources (RES), and electricity prices \cite{hon:etal:20:OAJPE,weron_electricity_2014}. However, as recently shown by Uniejewski and Weron \cite{uni:wer:21}, decisions based on probabilistic price forecasts, i.e., quantiles, prediction intervals or whole predictive distributions, can yield significantly higher profits. For the quantile-based bidding strategies considered in the Polish day-ahead power market, the profit obtained was from 5\% to 19\% higher than for the strategy based on point forecasts alone.

Point forecasts are far more popular in the electricity price forecasting (EPF) literature, not only in a decision support context. As reported by Maciejowska et al.\ \cite{mac:uni:wer:23}, probabilistic EPF was not part of the mainstream literature until the Global Energy Forecasting Competition in 2014 \cite{hong_probabilistic_2016}, and even now, no more than 15\% of the Scopus-indexed articles concern it. Although business analysts have begun to recognize their importance in the planning and operation of energy systems \cite[see e.g.][]{janczura_dynamic_2022}, it is not easy to generate accurate probabilistic predictions.
Combining forecasts obtained from different model specifications \cite{makridakis_m4_2018} or calibration sample lengths \cite{hubicka_note_2019,uni:mac:23} can significantly increase accuracy without sacrificing computational complexity or interpretability. Compared to selecting a single best-performing forecast, combining forecasts from multiple models offers several advantages, such as increased resilience against model uncertainty or misspecification, and better adaptability in the event of structural breaks \cite{wang_forecast_2022}.

Forecast combinations \chg{(also called} ensemble forecasts\chg{)} involve assigning weights to the individual predictions (or experts). While naive\chg{, i.e., equal,}  weighting is a straightforward -- and surprisingly robust -- way of averaging point forecasts, in the case of predictive distributions, a choice must be made about what to combine. Two natural approaches are vertical averaging of probabilities and horizontal averaging of quantiles \cite{lic:g-c:win:13,marcjasz_distributional_2023}, but the literature does not agree on which is better. Berrisch and Ziel \cite{berrisch_crps_2021} have recently proposed a cutting-edge weighting technique, called ``CRPS learning'', that accounts for variations in predictive performance over time and across quantiles of the distribution. It optimizes weights with respect to the continuous ranked probability score (CRPS), the standard error metric for probabilistic forecasts \cite{gneiting_strictly_2007,mac:uni:wer:23}.

In this article we address the question of whether forecast combinations obtained by minimizing the CRPS lead to optimal decisions in day-ahead bidding. To this end, we conduct a comprehensive empirical study involving: 
\begin{itemize}
    \item six years of hourly day-ahead electricity prices from the German EPEX market, 
    \item state-of-the-art probabilistic forecasts generated by Marcjasz et al.\ \cite{marcjasz_distributional_2023} using distributional deep neural networks (DDNN), as well as deep neural networks (DNN) and LASSO-estimated autoregressive (LEAR) models combined with quantile regression (QR),
    \item and two approaches to combining predictive distributions -- horizontal averaging of quantiles \cite{lic:g-c:win:13} and CRPS learning \cite{berrisch_crps_2021}.
\end{itemize}
Since statistical measures of forecast accuracy do not assess the utility of a forecast to its potential end users \cite{hon:etal:20:OAJPE,janczura_dynamic_2022,mac:uni:wer:23,yardley_error_2021}, we calculate the profits of a day-ahead bidding strategy  \cite{marcjasz_distributional_2023,uniejewski_smoothing_2023}. The latter aims to find the most financially beneficial hours of the next day to buy electricity and charge a battery, then discharge it and sell electricity. To minimize the risk of losses, limit orders are submitted to the power exchange with the limits determined by selected quantiles of the predictive distributions.

The remainder of the article is organized as follows. In Section \ref{sec:data} the dataset and assumptions for the forecasting problem are introduced. Section \ref{sec:methods} describes the details of ensemble construction. The results are presented in Section \ref{sec:evaluation}, with the forecast accuracy being the focus of Section \ref{ssec:accuracy}, the trading simulation described in \ref{ssec:battery} and its financial results in Section \ref{ssec:trading:profits}. Finally, Section \ref{sec:conclusions} wraps up the results and concludes.

\section{Preliminaries and data sources}\label{sec:data}

We assume a standard short-term forecast horizon of 1 day, performed in a rolling window scheme \cite{weron_electricity_2014}. More precisely, we assume that forecasts of all 24 hourly prices on day $d$ are calculated \chg{at the same time} in the morning of day $d-1$, i.e., before the day-ahead market for day $d$ closes, and that the model parameters are estimated using a calibration sample of $D$ most recent past observations. In our case, the underlying data are hourly day-ahead electricity prices from the German EPEX market spanning the period from 1 January 2015 to 31 December 2020. The prices as well as the day-ahead predictions of the loads and RES generation are publicly available from the ENTSO-E Transparency platform (\url{https://transparency.entsoe.eu}). The full dataset, including emission allowances and fuel prices, is also  available from \url{https://github.com/gmarcjasz/distributionalnn}, a GitHub repository that accompanies \cite{marcjasz_distributional_2023}. 

The first \chg{$D=1456$} days of the dataset are used as the initial calibration sample for all models, and additional 182 days are needed to calculate the quantile regression forecasts. The remaining 554 days from 27 June 2019 until 31 December 2020 constitute the out-of-sample test period. Note, that the latter includes a major drop in the level of prices associated with a decrease in the demand for electricity during the initial stage of the COVID-19 pandemic.

Since this study focuses on the evaluation of combination schemes for probabilistic forecasts and not on the computation of predictive distributions themselves, we work directly with a pool of readily available state-of-the-art forecasts generated by Marcjasz et al.\ \cite{marcjasz_distributional_2023}. The latter take the form of 99 predicted percentiles for each day and hour, which approximate the predictive distribution quite well. They are generated by twelve different models, eight of which are distributional deep neural networks (DDNN) with the output layer returning fitted parameters of the normal or Johnson's SU (JSU) distributions. Since the quantile functions have no closed form representations, the percentile forecasts are obtained as empirical quantiles of a 10000-element random sample generated from the output normal or JSU distribution. 
These forecasts are denoted further in the text as \m{DDNN\_N\_\{1--4\}} and \m{DDNN\_JSU\_\{1--4\}}, respectively, with the numbers representing the hyperparameter set used for tuning the DDNNs.

The remaining models directly predict 99 percentiles with the use of quantile regression averaging  \cite[QRA;][]{nowotarski_computing_2015} or quantile regression machine \cite[QRM;][]{marcjasz_probabilistic_2020} methods, applied to point forecasts of two well-performing benchmarks -- LASSO-estimated autoregressive models (LEAR) and deep neural networks (DNN) \cite{lag:mar:des:wer:21}. The combinations of these techniques make up the final four forecasts used in the ensembles: \m{LEAR\_QRA}, \m{LEAR\_QRM}, \m{DNN\_QRA} and \m{DNN\_QRM}. \chg{Note that in the LEAR models the prices for each of the 24 hourly load periods are treated as separate time series and estimated independently, whereas in the DNN and DDNN neural networks the 24 prices or 24 distributions are estimated jointly.} 

\section{Methods}\label{sec:methods}

Combining forecasts has become a well-established method to increase predictive accuracy. The advantages of using ensembles of experts in place of individual models include diversification of used information and increasing robustness against model misspecification and structural breaks in the data \cite{timmermann_chapter_2006}. 
While the literature generally recommends combining forecasts, many questions still remain open regarding the construction of ensembles. Across a multitude of possible specifications, the forecaster must decide on how many predictions to combine, how to perform forecast selection and how to choose weights for each expert. Combining probabilistic forecasts is even more tricky, as the assigned weights may change not just across experts and time, but also across quantile levels \cite{wang_forecast_2022}.

\subsection{Equal weighting}

In point forecasting, the use of naive\chg{, i.e., equal,} weights is often found to outperform more sophisticated weighting schemes because the latter introduce excessive estimation bias \cite{blanc_bias_2020}. In the case of predictive distributions, however, a choice must be made about what to combine. Two natural approaches are vertical averaging of probabilities, which boils down to computing a mixture distribution, and horizontal averaging of quantiles, where  each quantile of the ensemble forecast is a weighted average of the corresponding quantiles of all individual experts \cite{lic:g-c:win:13}. While the literature does not agree on which approach is better, Marcjasz et al.\ \cite{marcjasz_distributional_2023} emphasize that horizontal averaging is more robust and results in a sharper, i.e., more concentrated, unimodal distribution. On the other hand,  vertical averaging may lead to increased variance and multimodality. For this reason, as well as potential information loss due to interpolation needed to perform vertical averaging, only horizontal averaging of quantiles is considered in this paper. For consistency with other EPF studies, we denote it in the text by \m{qEns}.

\subsection{CRPS learning}

Berrisch and Ziel \cite{berrisch_crps_2021,berrisch_multivariate_2023} have recently proposed a cutting-edge weighting technique that accounts for variations in predictive performance over time and across quantiles; it is freely available in the \textit{profoc} package for R (\chg{\cite{profoc_package},} \url{https://cran.r-project.org/web/packages/profoc}). The authors called it ``CRPS learning'', since it optimizes weights with respect to the continuous ranked probability score (CRPS). The latter is a proper scoring rule and the standard error metric for probabilistic forecasts \cite{gne:kat:14,gneiting_strictly_2007}. It is  defined as: 
\begin{equation}
    \operatorname{CRPS}(F, x) = -\int_{-\infty}^{\infty} \left(F(y) - \mathbbm{1}_{\lbrace y \geq x \rbrace} \right)^2 dy,
\end{equation}
where $F$ is the cumulative distribution function of the evaluated probabilistic forecast. It can equivalently be represented as a scaled integral of the quantile loss, which for an equidistant grid can be approximated by:
\begin{equation}\label{eqn:CRPS:approx}
    \operatorname{CRPS}(F, x) \approx \frac{2}{M} \sum_{i=1}^M \operatorname{QL}_{p_i} \left(F^{-1}(p_i), x\right),
\end{equation}
where 
$\left(p_1, \ldots, p_M\right)$ 
is an equidistant monotonically increasing dense grid of probabilities and 
$\operatorname{QL}_p(q, x) = \left(\mathbbm{1}_{\lbrace x < q \rbrace} - p \right)\left(q - x\right)$
is the quantile loss for a quantile forecast $q$ of true value $x$ for probability $p \in (0,\ 1)$, also known as the pinball score \citep{berrisch_crps_2021,mac:uni:wer:23}.

The CRPS learning algorithm aims to combine probabilistic forecasts by selecting optimal weights for averaging across quantiles to minimize the CRPS of the resulting ensemble. The weight functions are subject to online updating throughout the forecasting period and are chosen pointwise, i.e., for each quantile of the distribution separately, depending on each \chg{expert's} performance. The framework additionally includes smoothing procedures which reduce estimation noise of the weights \cite{berrisch_multivariate_2023}.

In this study, the CRPS learning framework was applied once per ensemble, with the following arbitrarily chosen set of parameters: Bernstein online aggregation (BOA) for updating weights, penalized probabilistic smoothing with $\lambda = 2^{\left(-5, \ldots, 5\right)}$ updated based on past performance, and no forgetting past regret. The remaining options were set to the \textit{profoc} package defaults. Such an approach is denoted in the text by \m{CRPS}. Finally, note that the time required to compute forecasts of a single CRPS learning ensemble for the entire test period is ca.\ 500 times longer than for the naive \m{qEns} weighting. However, it does not exceed 20~s on a laptop equipped with a 9th generation Intel Core i7-9750H processor.

\begin{figure}
    \centering
    \includegraphics[width=\textwidth]{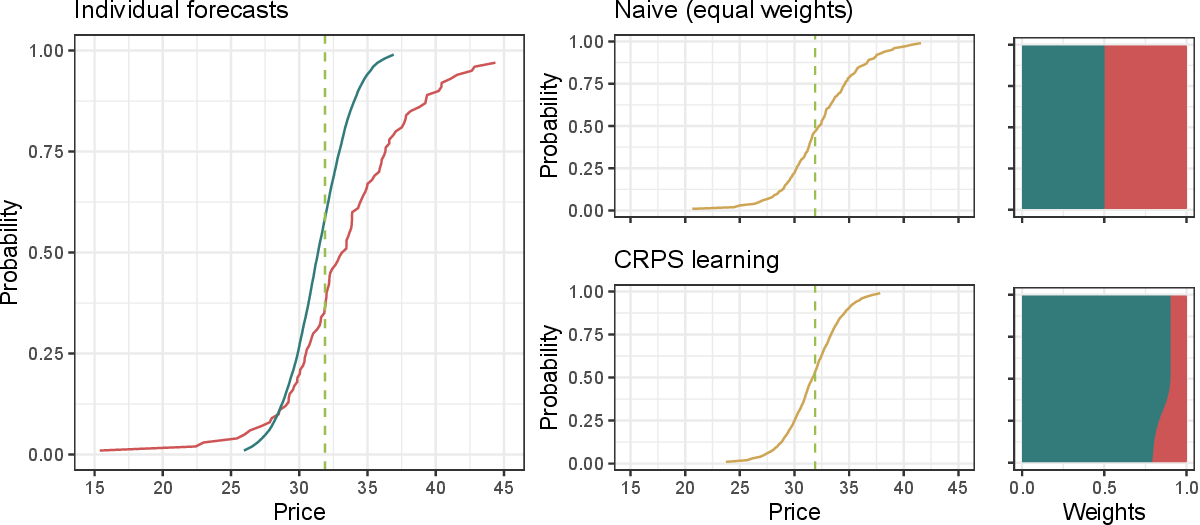}
    \caption{Illustration of the two weighting schemes. The left panel shows predictive distributions obtained from the \m{DDNN\_JSU\_1} (teal color) and \m{LEAR\_QRA} (red color) models for a selected hour and day. The center panels present the resulting ensemble forecasts obtained by estimating weights with naive (top) and CRPS learning (bottom) methods. The right panels illustrate the relative weights for each quantile; these are horizontal stacked bar plots with the length of the bar representing the weight of the forecast in corresponding color and all weights summing up to 1. The dashed vertical line marks the actual price.}
    \label{fig:forecast_comb}
\end{figure}

\subsection{Comparison of the two weighting schemes}

The general idea of averaging across quantiles, as well as differences between the two weighting schemes, are shown in Figure \ref{fig:forecast_comb}. The illustration shows a toy example of a two-forecast ensemble. Among the two experts, the \m{DDNN\_JSU\_1} forecast (teal color) is sharper, i.e., more concentrated, predicting prices between 26 and 37~EUR, and has a smoother cumulative distribution function (CDF), while the \m{LEAR\_QRA} predictive distribution (red color) is less  sharp (with prices between 15 and 44~EUR) and more rugged. Medians of both experts are relatively close to the actual observed price (\chg{31.89~EUR,} vertical line), albeit leaving room for improvement\chg{, i.e., with absolute errors of 0.52 EUR and 1.24 EUR, respectively}.

The two individual forecasts are combined using the two weighting approaches, with the resulting CDFs and the assigned weights shown in the panels to the right. It can be seen that while the \m{qEns} approach, by definition, assigns equal weights to all models and quantiles, CRPS learning assigns larger weights to the \m{DDNN\_JSU\_1} forecast, based on its better past performance (not shown in the plot). The share of the \m{DDNN\_JSU\_1} forecast is smaller for the lowest 25 percentiles, but nevertheless it still dominates the \m{CRPS} ensemble, leading to its higher sharpness (price range of [24, 38] EUR) and smoothness compared to the equally weighted ensemble (with values in the range [21, 42] EUR). However, both forecast combinations provide a more accurate median forecast than the individual experts\chg{, with absolute errors of 0.36~EUR for \m{qEns} ensemble and 0.22~EUR for CRPS learning}.

\subsection{Selection of experts}

\begin{figure}
    \centering
     

\begin{tikzpicture}[
bignode/.style={rectangle, draw=black, fill=black!5, minimum size=20mm, minimum height=40mm, text centered},
smallnode/.style={rectangle, draw=black, fill=black!5, minimum size=20mm, minimum height=15mm, text centered},
every text node part/.style={align=center},
]

\draw[draw=black,fill=blue!5] (1.5, -2.25) rectangle (7.5, 2.25);
\node at (4.5,0) {(Marcjasz et al. \cite{marcjasz_distributional_2023})};

\node[bignode] (data) {Calibration \\ sample  \\ (Sec. 2)};
\node[smallnode] (pointforecasts) [right=of data.south east, anchor=south west] {Point \\ forecasts};
\node[smallnode] (qrforecasts) [right=of pointforecasts] {Quantile \\ regression \\ forecasts};
\node[smallnode] (nnforecasts) [above=of qrforecasts] {DDNN \\ forecasts};
\node[bignode] (experts) [right=of nnforecasts.north east, anchor=north west ] {Sets of \\ experts \\ (Sec. 3.4)};
\node[bignode] (ensembles) [right=of experts] {Weighted \\ ensemble \\ forecasts \\ (Sec. 3.1--3.3)};

\draw[thick, -Stealth] (1, 1.25) -- (nnforecasts);
\draw[thick, -Stealth] (1, -1.25) -- (pointforecasts);
\draw[thick, -Stealth] (pointforecasts) -- (qrforecasts);
\draw[thick, -Stealth] (nnforecasts) -- (8, 1.25);
\draw[thick, -Stealth] (qrforecasts) -- (8, -1.25);
\draw[thick, -Stealth] (experts) -- (ensembles);

\end{tikzpicture}
    \caption{\chg{Schematic illustration of the process of generating ensemble forecasts.}}
    \label{fig:schematic}
\end{figure}
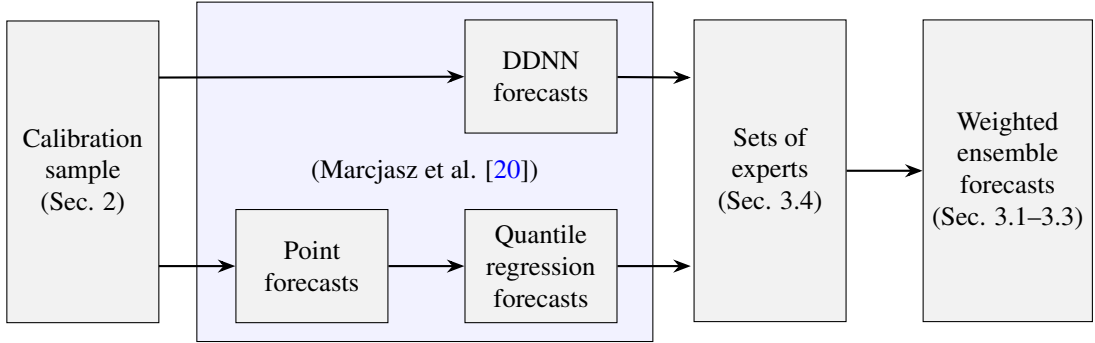

The second decision to be made by the forecaster is the selection of experts that are aggregated in the ensemble. Following \cite{marcjasz_distributional_2023}, each ensemble we consider in this study contains a set of four DDNN forecasts, either \m{DDNN\_N\_\{1--4\}} or \m{DDNN\_JSU\_\{1--4\}}. Furthermore, to diversify the pool of experts we additionally include benchmark quantile regression-based forecasts -- either \m{LEAR\_QRA} and \m{LEAR\_QRM} or \m{DNN\_QRA} and \m{DNN\_QRM}. While they have been demonstrated to perform significantly worse on their own, using them can lead to a higher prediction accuracy of the ensembles by avoiding overfitting \cite{wang_forecast_2022}.
Thus, the resulting naming convention for ensembles is:
$$
\m{DDNN\_\{distribution\}\_\{averaging\}\_\{experts\}},
$$
with \m{distribution=\{N,JSU\}}  denoting whether normal or JSU forecasts were used, \m{averaging=\{qEns, CRPS\}} indicating the use of equal or CRPS learning-derived weights, and \m{experts=\{LEAR,DNN\}} added when additional experts -- respectively \m{LEAR\_QRA} and \m{LEAR\_QRM} or \m{DNN\_QRA} and \m{DNN\_QRM} -- were included in the ensemble. \chg{A graphical illustration of the steps performed in order to construct the ensemble forecasts is shown in Figure \ref{fig:schematic}.}

It should be noted that additional forecast combinations were explored during the course of this research. The complete list included smaller ensembles (four DDNN forecasts and a single QR-based forecast; best 5 performing models), other combinations of quantile regression forecasts (e.g., four DDNN forecasts, \m{LEAR\_QRA}, \m{DNN\_QRA}) and larger ensembles (four DDNN forecasts and four QR-based forecasts; eight DDNN forecasts as in \cite{berrisch_multivariate_2023}; all available forecasts). 
They offered comparable or, in the case of the largest ensembles, significantly inferior performance to the combinations listed above, and have been omitted from the presentation of results for the sake of clarity.

\chg{In an online learning setting, the forecaster may decide to discard the initial part of the test sample as a burn-in period, which is beneficial for the stability of weights and hyperparameters, see, e.g., \cite{berrisch_multivariate_2023}, which uses a burn-in period of 182 days for combinations of DDNN forecasts. However, the quantile regression forecasts are only available within the 554-day out-of-sample test period, the entirety of which has been used in Berrisch et al.\ \cite{berrisch_multivariate_2023} and Marcjasz et al.\ \cite{marcjasz_distributional_2023} for evaluation. Since the majority of ensembles we consider in this study include these quantile regression forecasts, for the sake of consistency no burn-in period has been applied.}

\section{Forecast evaluation}\label{sec:evaluation}

In this section, the generated ensemble forecasts are compared to each other and the individual expert models. The evaluation is divided into two parts. First, in Section \ref{ssec:accuracy}, we measure the predictive accuracy in terms of statistical error metrics:
\begin{itemize}
    \item the mean absolute error (MAE) and the root mean squared error (RMSE) for median and mean forecasts, respectively \chg{\cite{gneiting_making_2011}}, 
    \item the continuous ranked probability score (CRPS) for probabilistic forecasts \chg{\cite{gneiting_strictly_2007,nowotarski_recent_2018a}}.
\end{itemize}
Then, in Section \ref{ssec:battery}, we measure the predictive accuracy in terms of profits -- total and per trade -- from a day-ahead bidding strategy that utilizes probabilistic forecasts \cite{marcjasz_distributional_2023,uniejewski_smoothing_2023}.
Note, that the CRPS is approximated by a sum of pinball scores on a grid of 99 percentiles, see Eqn.\ \eqref{eqn:CRPS:approx}. The statistical significance of differences in CRPS scores is assessed using the Diebold-Mariano test \cite{diebold_comparing_1995}.

\subsection{Evaluation in terms of statistical error measures}\label{ssec:accuracy}

As Gneiting et al.\ \cite{gne:kat:14,gneiting_strictly_2007} argue, the goal of probabilistic forecasting is to ``maximize the sharpness of the predictive distributions subject to calibration''. Here, calibration (also called reliability or unbiasedness) refers to the statistical consistency between the probabilistic forecasts and the observations, e.g., whether the 50\% prediction interval (PI) covers 50\% of the actual observations. Sharpness, on the other hand, refers to the concentration of the predictive distributions. For instance, given two reliable 50\% PIs, the sharper or more narrow one is better. The CRPS introduced in Section \ref{sec:methods} assesses calibration and sharpness simultaneously \cite{gne:kat:14}. Moreover, for a point forecast, it is equal to the MAE
\cite{nowotarski_recent_2018a}.

\begin{figure}
    \centering
    \includegraphics[width=\textwidth]{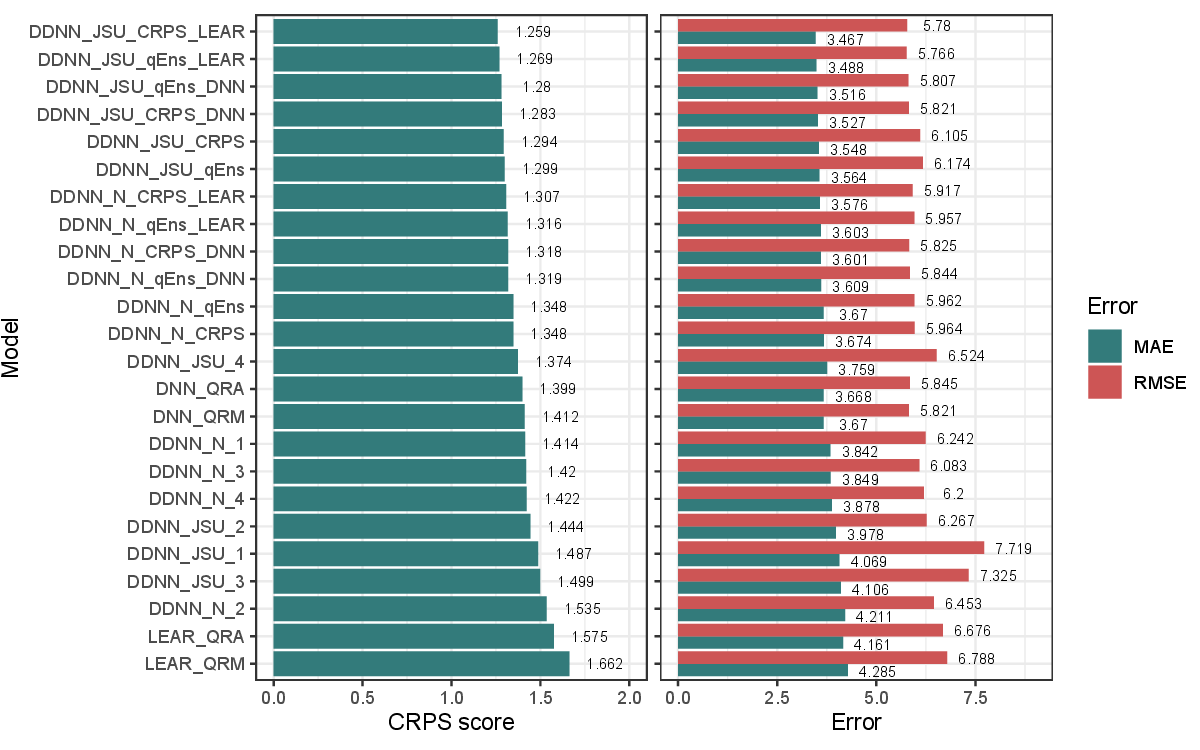}
    \caption{CRPS scores (left panel) and MAE and RMSE errors (right panel) for all models and ensembles, ordered from the lowest to the highest CRPS. Compare with a CRPS of 1.284 of the best performing model of Berrisch and Ziel \cite{berrisch_multivariate_2023}.}
    \label{fig:errors}
\end{figure}

The CRPS values for all models ordered from the lowest/best to the highest/worst are shown in the left panel of Figure \ref{fig:errors}; the corresponding MAE and RMSE errors in the right panel.
Clearly, the two \m{LEAR} forecasts perform the worst, while the \m{DDNN\_JSU\_4} and two \m{DNN} forecasts the best out of the individual models. Moreover, the individual experts are outclassed by all \m{DDNN\_N} ensembles, which are further outperformed by the \m{DDNN\_JSU} combinations. The \m{DDNN\_JSU\_CRPS\_LEAR} ensemble achieves the lowest CRPS score. For all ensembles, both weighting schemes typically result in very similar forecasts and thus accuracy. Nevertheless, CRPS learning yields slightly better predictions on average.
A similar, though not identical, ordering can be observed for the point forecasting error metrics. The most significant differences are obtained for the \m{DDNN\_JSU\_\{1,3,4\}} experts in terms of the RMSE. 
As in \cite{marcjasz_distributional_2023}, we calculate the MAE for the median (i.e., the 50th percentile) and the RMSE for the expected value of each distribution; 
the errors presented for the individual forecasts as well as the \m{DDNN\_N\_qEns} and \m{DDNN\_JSU\_qEns} ensembles are consistent with the results reported in \cite{marcjasz_distributional_2023}. 

\begin{figure}
    \centering
    \includegraphics[width=0.9\textwidth]{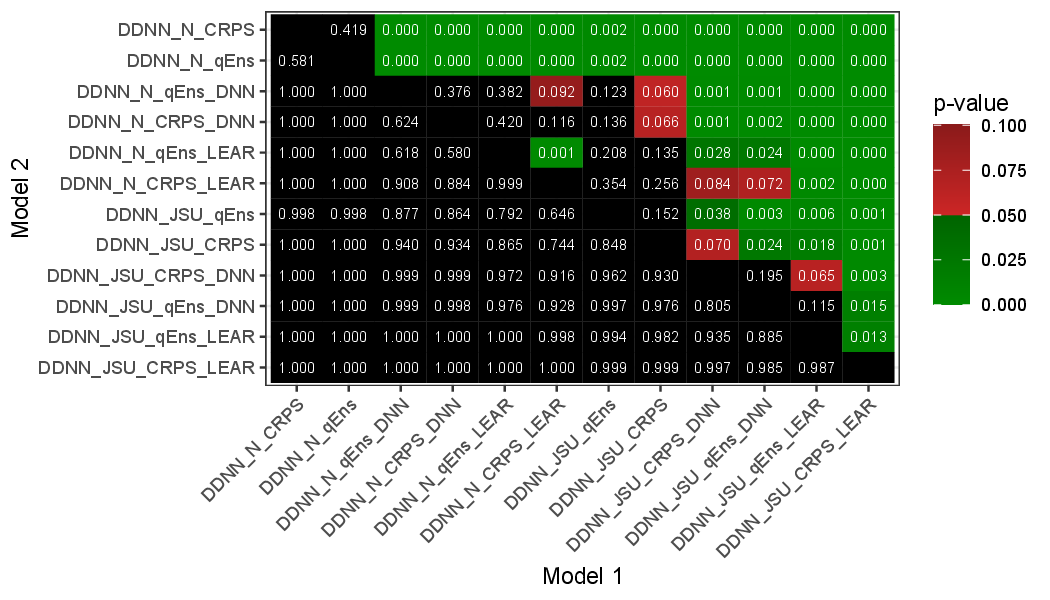}
    \caption{Results ($p$-values) of the Diebold-Mariano test for the CRPS loss -- the lower it is the more significant is the difference between the forecasts of a model on the X-axis (better) and the forecasts of a model on the Y-axis (worse). We use a coloring scheme to highlight the differences.}
    \label{fig:DM_CRPS}
\end{figure}

To assess statistical significance of differences in CRPS scores we perform the Diebold-Mariano (DM) test \citep{diebold_comparing_1995}. In order to correct for daily seasonality in CRPS values, following \cite{lag:mar:des:wer:21} and \cite{marcjasz_distributional_2023}, we consider a multivariate loss differential series defined for a pair of models $A$ and $B$ as:
\begin{equation}
    \Delta_{d}^{A, B} = \|L_d^A\|_1 - \|L_d^B\|_1,
\end{equation}
where $L_d^X = \{L_{d,1}^X,\ldots,L_{d,24}^X\}$ is the 24-dimensional vector of hourly CRPS values for model $X$ on day $d$ and $\|L_d^X\|_1$ is its $L_1$ norm. For each pair of models we apply two one-sided DM tests.

A heatmap of the respective $p$-values is presented in Figure \ref{fig:DM_CRPS}. The results indicate that the predictions of the \m{DDNN\_JSU\_CRPS\_LEAR} ensemble are significantly better than those of all competing models. The predictions of the remaining ensembles within the top 4 do not significantly differ among each other. Another ensemble whose forecasts are significantly better than those ranked lower in terms of the CRPS is \m{DDNN\_JSU\_CRPS\_DNN}, while most other ensembles do not yield significantly better predictions than ensembles similarly ranked in terms of the CRPS.

The CRPS score provides a single number for all quantiles (and \chg{each time point} in the test period). To see how the pinball scores for individual percentiles contribute to the CRPS, in Figure \ref{fig:pinball_check} we plot them for selected best performing ensembles. To enhance readability, all values are plotted with respect to the pinball scores of the best performing ensemble, i.e., \m{DDNN\_JSU\_CRPS\_LEAR}. Clearly, the relative performance of the ensembles is not uniform across the entire distributions. The largest disparity can be seen below the median, with the relative ranking of the ensembles changing for the three lowest percentiles. Above the median, the ensembles perform similarly.

\subsection{Day-ahead bidding}\label{ssec:battery}

Following Uniejewski \cite{uniejewski_smoothing_2023}, we consider a realistic trading strategy that utilizes battery storage and day-ahead bidding based on probabilistic price forecasts. The goal is to buy electricity cheaply at hour $h1$ and charge the battery, then discharge it and sell the electricity expensively at hour $h2>h1$. To minimize the risk of losses, limit orders are submitted to the power exchange with the limits determined by selected -- based on trader's risk appetite -- quantiles of the predictive distributions.

We assume that the efficiency of charging as well as discharging the battery is 90\%. Hence, $\frac{1}{0.9}\approx 1.1$ MWh is needed to charge the battery by 1 MWh. Similarly, discharging 1 MWh generates only 0.9 MWh. Further, we assume that the total usable capacity of the battery is $B=2$ MWh and that at the beginning of the simulation period the battery starts halfway charged  ($B=1$). If both orders are executed on the next day, this state persists. 
If $B=0$ at the beginning of a day, an unlimited bid to buy 1 MWh is placed at hour $h*<h2$, and if $B=2$, an unlimited offer to sell 1 MWh is placed at hour $h*<h1$. 

For each day in the out-of-sample test period, the following two steps are performed. First, based on median price forecasts $Y^{0.5}_{d,h}$ for day $d$ and hours $h=1,2,...,24$ computed on day $d-1$, hours $h1$, $h2$ and $h*$ are selected to maximize the profit:
\begin{equation}\label{eq:profit}
    \Pi_d = -\frac{1}{0.9} \hat{Y}^{0.5}_{d,h1} 
    + 0.9 \hat{Y}^{0.5}_{d,h2} 
    - \mathbbm{1}_{\lbrace B=0 \rbrace} \frac{1}{0.9} \hat{Y}^{0.5}_{d, h*}
    + \mathbbm{1}_{\lbrace B=2 \rbrace} 0.9 \hat{Y}^{0.5}_{d, h*}.
\end{equation}
When the battery is halfway charged ($B=1$) this optimization problem reduces to selecting hours with the lowest and the highest predicted median price. In other cases, linear programming is used to optimize the selection of $h1$, $h2$ and $h*$.

\begin{figure}
\centering
    \begin{minipage}[t]{0.49\textwidth}
    \includegraphics[width=\textwidth]{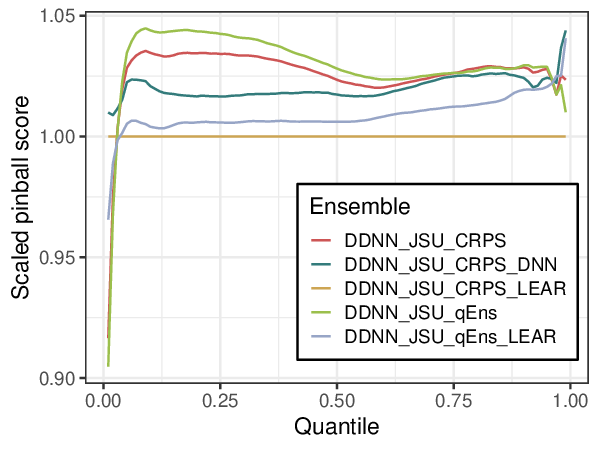}
    \caption{Pinball scores of selected best performing ensembles across quantiles, relative to the \m{DDNN\_JSU\_CRPS\_LEAR} ensemble. A lower score corresponds to better performance.}
    \label{fig:pinball_check}
    \end{minipage}
    ~
    \begin{minipage}[t]{0.49\textwidth}
    \includegraphics[width=\textwidth]{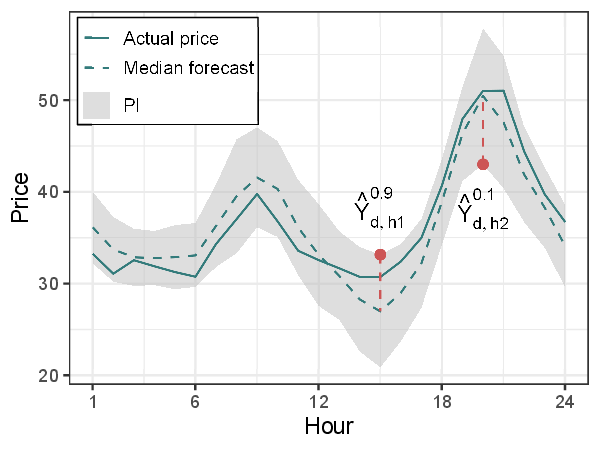}
    \caption{Illustration of the trading strategy with limit orders defined by the 80\% PIs, corresponding to a risk appetite of 0.8. Red dots indicate the price limits for the selected hours. 
    }
    \label{fig:graphic_strategy}
    \end{minipage}
\end{figure}

Next, following Marcjasz et al.\ \cite{marcjasz_distributional_2023}, a profitability condition is checked. If the transaction is expected to be profitable, i.e., the sum of the first two terms in Eqn.\ \eqref{eq:profit} is greater than zero, a buy order with price limit $\hat{Y}_{d,h1}^{1-q}$ and a sell order with price limit $\hat{Y}_{d,h2}^{q}$ are placed. Here $q=\frac{1-\alpha}{2}$ and $\alpha$ is trader's risk appetite, i.e., the PI level, set only once for the whole test period. This is illustrated in Figure \ref{fig:graphic_strategy} for a sample day and forecasts generated by the \m{DDNN\_N\_qEns} ensemble. In this example both orders would be accepted since the actual price falls within the 80\% PI, corresponding to a risk appetite of $\alpha=0.8$. However, hour $h_2$ is predicted suboptimally, a slightly higher price was observed for hour 21.

\subsection{Evaluation in terms of trading profits}
\label{ssec:trading:profits}

The total profits are presented in Table \ref{tab:total_income_heatmap} for five values of risk appetite $\alpha$ ranging from 0.5 to 0.9; the minor differences between the reported values and those in \cite{marcjasz_distributional_2023} for the \m{DDNN\_N\_qEns} and \m{DDNN\_JSU\_qEns} ensembles are a result of correcting a  bug in the original software.
The profitability results somewhat correspond to the CRPS results, although with a few notable exceptions. On average, the \m{DDNN\_JSU} ensembles achieve higher total profits than the \m{DDNN\_N} ensembles, mirroring their better performance in terms of the CRPS. The detailed ranking of those ensembles is where the outcomes start to vary. While the CRPS weighting scheme outperforms equal weights in terms of forecast accuracy, this trend is mostly reversed in the financial results. This is especially true for lower values of the risk appetite. 

\begin{figure}[p]
    \centering
        \includegraphics[width=\textwidth]{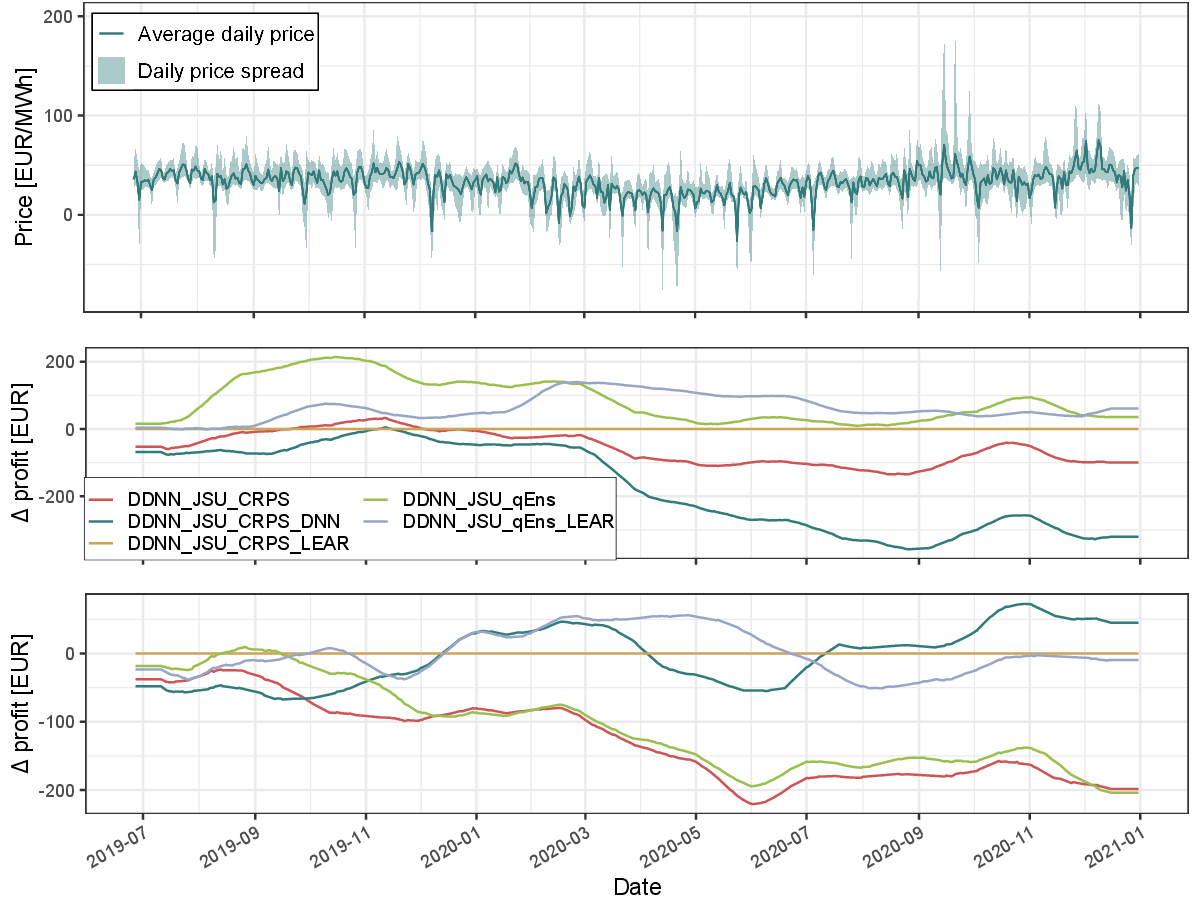}
        \caption{Average daily (dark green) and the minimum and maximum hourly (light green) prices in Germany from 27 June 2019 to 31 December 2020 (top panel). 30-day moving average of cumulative profit for the best performing strategies, shown as a difference between cumulative profit of each ensemble and \m{DDNN\_JSU\_CRPS\_LEAR}, for risk appetite $\alpha=0.6$ (center panel) and $\alpha=0.8$ (bottom panel).}
        \label{fig:income_in_time}
\end{figure}

For instance, the most accurate in terms of the CRPS ensemble, i.e., \m{DDNN\_JSU\_CRPS\_LEAR}, yields lower profits than its \m{qEns} counterpart. This is likely due to the fact that while the CRPS weighting is more accurate on average, it is significantly outperformed by the naive weighting for the few lowest percentiles, see Figure \ref{fig:pinball_check}, giving the latter an advantage during the initial stage of the COVID-19 pandemic (middle part of the test period), see Figure \ref{fig:income_in_time}. A similar behavior can be observed for the \m{DDNN\_JSU\_CRPS\_DNN} ensemble, which performs worse than its competitors for the extreme quantiles, being at a significant disadvantage in the second half of the evaluation period, when the daily price spread is higher than in the beginning.

\begin{table}[p]
    \centering
    \caption{Total profits from the quantile-based trading strategy in the whole test period for risk appetite ranging from 0.5 to 0.9. The highest values in each column are in bold. Cells are colored independently in each column from the best ($\rightarrow$ green) to the worst ($\rightarrow$ red).}
    \includegraphics[width=\textwidth]{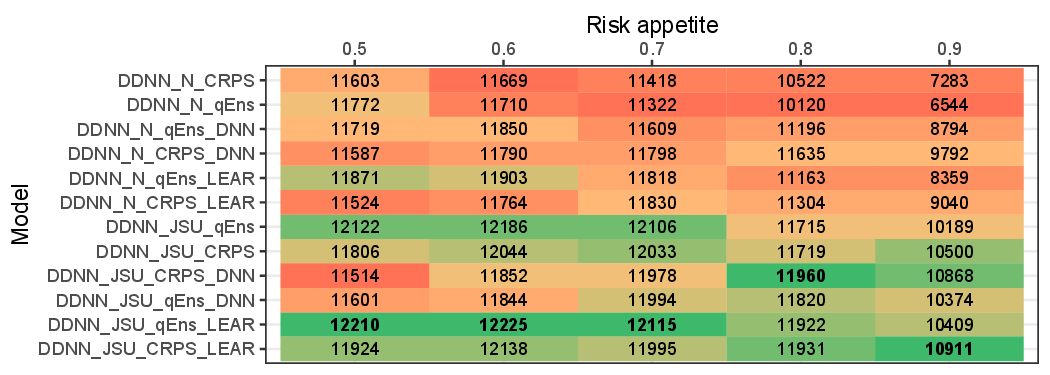}
    \label{tab:total_income_heatmap}
\end{table}

It can be expected that an optimal trading strategy would result in executing exactly two trades per day, buying on the low and selling on the high. With such a ``crystal ball'' strategy, the trader would earn 13587~EUR throughout the whole evaluation period. Conversely, taking the worst possible decisions would lead to a total loss of $-21425$~EUR. On this scale of possible profits, all evaluated ensembles rank relatively well. The lowest profit presented in Table \ref{tab:total_income_heatmap} reaches 80\% of the maximum, while the best of all forecasts as much as 96\%. For comparison, a naive strategy of placing bids at fixed hours  selected \emph{ex-post} as having the highest price spread on average -- buying at hour 3 and selling at hour 19 -- would lead to total profits of 8048~EUR, or 84\% of the maximum, see \cite{marcjasz_distributional_2023}.

The profit per trade results reported in Table \ref{tab:income_per_trade_heatmap} are less clear-cut, with the \m{DDNN\_N} ensembles no longer being completely outclassed by the \m{DDNN\_JSU} ensembles, especially when there are fewer total trades. As profits per trade can be seen as an indicator of the trader's risk, this disparity is consistent with literature findings \cite{janczura_arxgarch_2023}.
It is worth noticing that, perhaps unintuitively, higher values of the risk appetite correspond to higher risk aversion. This is an effect of the final step of the strategy, which checks the income of the worst case scenario. With higher values of the risk appetite, this predicted profit tends to be lower, leading the trader to act more cautiously. This seemingly leads to dominance of models which are more accurate across the entire distribution rather than only in the extreme quantiles,  compare the center (risk appetite $\alpha=0.6$) and bottom (risk appetite $\alpha=0.8$) panels of Figure \ref{fig:income_in_time} with Figure \ref{fig:pinball_check}. However, this is only a conjecture, as the relationship between daily price levels, price spreads and PIs is not direct or linear.

\begin{table}
    \centering
    \caption{Profits per trade from the quantile-based trading strategy in the whole test period for risk appetite ranging from 0.5 to 0.9. The highest values in each column are in bold. Cells are colored independently in each column from the best ($\rightarrow$ green) to the worst ($\rightarrow$ red).}
    \includegraphics[width=\textwidth]{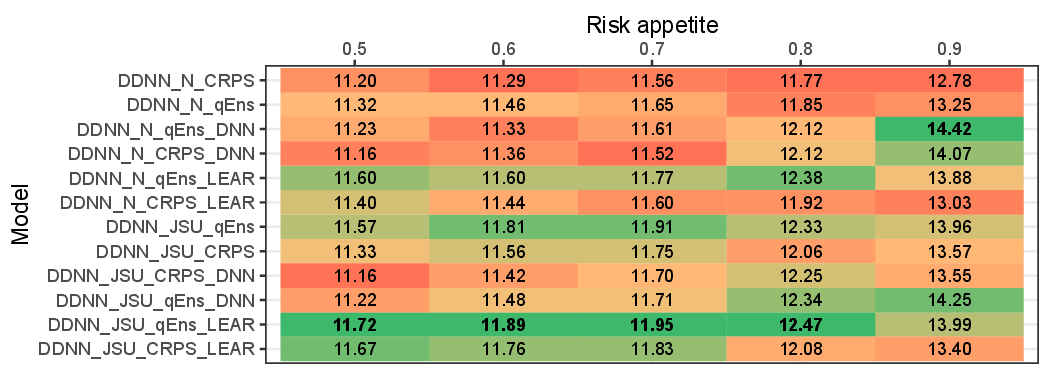}
    \label{tab:income_per_trade_heatmap}
\end{table}

\section{Conclusions and discussion}\label{sec:conclusions}

In this article we address the question of whether minimizing the continuous ranked probability score (CRPS) -- the standard error metric for probabilistic forecasts -- leads to optimal decisions in day-ahead bidding. Conducting an extensive empirical study, we find that introducing diversity to a  pool of forecasts is highly beneficial, both in terms of forecast accuracy measured by the CRPS and profits from a trading strategy implemented in the German day-ahead power market. Also optimizing combination weights with CRPS learning positively impacts forecast accuracy. This is likely caused by the uneven performance of experts across time and quantiles, which is an outcome consistent with the literature.

While trading profits generally follow forecast accuracy, the benefits of using CRPS learning are not as pronounced in the trading scenario, especially considering the ca.\ 500 times higher computational burden. The precise cause-and-effect relationships between the predictive accuracy and profits are difficult to disentangle. However, the performance for the extreme quantiles of the distribution seems to be related to some of the observed patterns. In general, using any of the considered ensembles leads to achieving satisfactory profits, especially when compared to the best-case and worst-case scenarios.

For the sake of clarity, only selected forecast sets were considered. Extending the pool of experts and ensembles could lead to a more comprehensive evaluation. Other possible extensions of this study include comparisons with other weighting schemes, as well as automated methods for expert selection.

\acknowledg{This research was partially supported by the Ministry of Science and Higher Education (MNiSW, Poland) through Diamond Grant No. 0027/DIA/2020/49 (to W.N.) and the National Science Center (NCN, Poland) through grant No. 2018/30/A/HS4/00444 (to R.W.).}

\bibliographystyle{acm}
\bibliography{NitkaWeron23_ORD}

\end{document}